\documentclass[english,prb,twocolumn]{revtex4-1}
\usepackage[T1]{fontenc}
\usepackage[latin9]{inputenc}
\usepackage{amsmath}
\usepackage{amssymb}
\usepackage{graphicx}

\makeatletter

\newcommand{\noun}[1]{\textsc{#1}}

\makeatother

\usepackage{babel}
\begin{document}
\title{Effects of correlated magnetic noises from shared control lines on
two-qubit gate }
\author{Yinan Fang$^{1,2}$}
\email{ynfang@ynu.edu.cn}

\affiliation{$^{1}$School of Physics and Astronomy and Yunnan Key Laboratory for
Quantum Information, Yunnan University, Kunming 650500, China~\\
$^{2}$Beijing Computational Science Research Center, Beijing 100193,
People's Republic of China}
\begin{abstract}
Recent proposals for building scalable quantum computational devices
in semi-conductor based spin qubits introduce shared control lines
in order to reduce the overhead of qubits controls. In principle,
noises from the shared controls could introduce correlated errors
to multi-qubit gates, and identifying them will be helpful for achieving
higher gate fidelity in those setups. Here, we introduce a method
based on the randomized benchmarking protocols that is capable of
distinguishing among different correlated noises in a particular two-qubit
model motivated by the crossbar architecture. 
\end{abstract}
\maketitle

\section{Introduction}

Fault-tolerant quantum computation requires synergy over a large number
of physical qubits: Depending on the error rate, a logical qubit based
on the surface code can take thousands of physical qubits to encode.\citep{2012_PRA_Fowler,2018_Quantum_Preskill}
This motivates the construction of large scale quantum computational
devices. Despite that single-qubit and two-qubit gates with the gate
fidelity above the fault-tolerant threshold had been achieved across
various platforms,\citep{2016_PRL_Ballance,2022_Nature_Noiri,2022_Nature_Xue}
upto now only 72 qubits were realized experimentally based on the
superconducting circuit,\citep{2023_Nature_Acharya} and it was not
until recently that the logical qubit could be demonstrated.\citep{2022_PRL_Zhao,2022_Nature_Krinner}

One of the difficulties involved in building a desired large scale
device capable of implementing error correction code is to maintain
uniformity in the characteristic parameters across all the underlying
qubits. \citep{2021_Nanotehcnol_Laucht,2023_NanoLett_Meyer} In this
regard, semi-conductor based spin qubits stands out with a unique
advantage of being compatible with the matured industrial fabrication
technologies. \citep{2017_NatCommun_Veldhorst} Following this approach,
several proposals for building scalable devices had been put-forward.\citep{2015_SciAdv_HillCharles,2017_NatCommun_Veldhorst,2018_SciAdv_Li,2021_NatureElectron_GonzalezZalba}
Later in experiments, a two-dimensional crossbar array containing
$36\times36$ gates were reported,\citep{2022_npjQuantum_Bavdaz}
while recently a two-dimensional array of 16 qubits with the inter-dot
coupling above 10GHz was experimentally realized in terms of the SiGe
based hole-spin quantum dots. \citep{Borsoi2023} 

Unlike the traditional few-qubit setups for the proof-of-principle
demonstrations where each qubit could be individually addressable,
those scalable proposals usually features common control lines that
borrowed from the CMOS or crossbar architectures.\citep{2017_NatCommun_Veldhorst,2018_SciAdv_Li}
On one hand they reduce the complexity of control layers and enable
the integration of plenty of qubits, on the other hand, however, the
noises imposed on such shared control lines may affect several qubits
at the same time, thus result in \emph{correlated} errors among the
qubits. Although cryo-CMOS techniques have been developed for the
purpose of qubit operation,\citep{2018_IEEE_Patra,2021_IEEE_Charbon}
since those noises differ from the influences of the local environment
that experienced by individual qubit, it would be still useful to
distinguish between those noises in order to further reduce the control
errors that associated with the multi-qubit operations.

In this work, we explore several possible sources of correlated magnetic
noises in a simplified two-qubit model motivated by the crossbar proposal
for scalable quantum computation. \citep{2018_SciAdv_Li} To distinguish
among those noises, we apply a method based on the interleaved randomized
benchmarking (IRB). \citep{2012_PRL_Magesan} In additional to the
standard random circuit in the IRB, we introduce an additional measurement
induced decoherence process to selective address certain components
of the noisy quantum channel. Although the investigation is based
on a specific model, the method should be generally applicable also
to other scenarios.

This paper is organized as follows: In Sec. II, we introduce the model
and derive the magnetic field at the qubit location. In Sec. III,
different possible magnetic noises are discussed, with the corresponding
perturbative Hamiltonian given explicitly. In Sec. IV, the method
based on the IRB is introduced and applied to distinguish between
two particular correlated magnetic noises. Further discussions are
provided in Sec. V, and we draw the conclusion in Sec. VI. Technique
details can be found in the Appendices.

\section{Magnetic field profile}

\begin{figure}
\begin{centering}
\includegraphics[width=8.6cm]{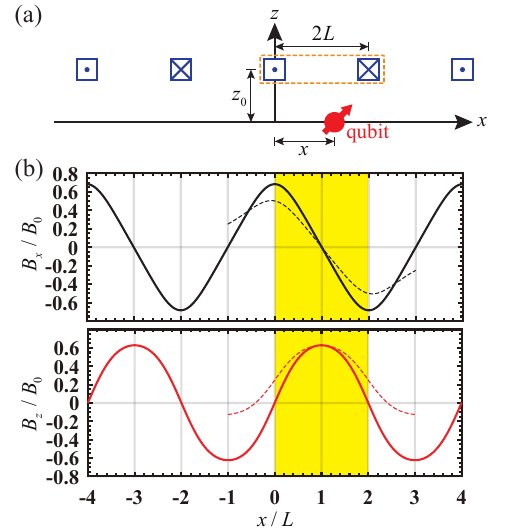}
\par\end{centering}
\caption{(a) Schematic plot of a spin qubit subjecting to magnetic field that
generated from parallel wires carrying alternating d.c. current. (b)
The total magnetic field as a function of the qubit location. Solid
curves are calculated from Eq. (\ref{Bfield}) with a cut-off $100$
for the $n$ summation, the dashed curves are obtained from the two
nearest wires {[}highlighted by the dashed orange box in (a){]} with
a calibration coefficient, cf. Eq. (\ref{Bfield_aprx}). In above
calculation we have assumed $z_{0}=L$ for clarity, while for $z_{0}\ll L$
the agreement between the solid and dashed curves will be improved.
\label{FIG_schematics}}
\end{figure}

Following several papers about the proposal of crossbar architecture
for salable quantum computation, \citep{2015_SciAdv_HillCharles,2017_NatCommun_Veldhorst,2018_SciAdv_Li}
here we consider the following simplified model that schematically
shown in FIG. \ref{FIG_schematics}(a). The system features an array
of equally spacing parallel wires carrying currents in alternating
directions. For an estimation of the magnetic field generated by the
$n$-th wire at the qubit position, we use an effective media description
such that \citep{1999_BOOK_Jackson}
\begin{equation}
\mathbf{b}_{n}=\frac{\mu_{\mathrm{eff}}I}{2\pi r_{n}^{2}}\mathbf{e}_{n}\times\mathbf{r}_{n},\label{SingleWireB}
\end{equation}
where $\mu_{\mathrm{eff}}$ is the effective 
 permeability of the media, $I$ is the current through the $n$-th
wire with the current direction indicated by the unit vector $\mathbf{e}_{n}$.
$\mathbf{r}_{n}$ is the position vector in the $x$-$z$ plane jointing
from the wire to the spin qubit. The total magnetic field experienced
by the qubit then amounts to the sum over the individual contributions
from each wire
\begin{equation}
\mathbf{B}=\sum_{n}\mathbf{b}_{n},
\end{equation}

Specifically, for a qubit located at $(x,0,0)$ under the configuration
of FIG. \ref{FIG_schematics}(a), the total magnetic field is given
by
\begin{equation}
\mathbf{B}=B_{x}\mathbf{e}_{x}+B_{z}\mathbf{e}_{z},\label{Bfield}
\end{equation}
\begin{equation}
B_{x}=B_{0}\sum_{n=-\infty}^{\infty}\left(\frac{z_{0}^{2}}{r_{n,1}^{2}}-\frac{z_{0}^{2}}{r_{n,2}^{2}}\right),
\end{equation}
and
\begin{equation}
B_{z}=B_{0}\sum_{n=-\infty}^{\infty}\left(\frac{z_{0}x_{n,1}}{r_{n,1}^{2}}-\frac{z_{0}x_{n,2}}{r_{n,2}^{2}}\right).
\end{equation}
Here, $B_{0}=\mu_{\mathrm{eff}}I/(2\pi z_{0})$ is the magnetic field
generated by a single wire provided that the qubit is located right
below the wire with the distance $z_{0}$. $r_{n,j}=\sqrt{x_{n,j}^{2}+z_{0}^{2}}$
and $x_{n,j}$ with $j=1,2$ are defined respectively as follows:
\begin{equation}
x_{n,1}=x-4Ln,\text{ }x_{n,2}=x-2L(2n+1).
\end{equation}

The magnetic field profile calculated according to Eq. (\ref{Bfield})
is shown in FIG. \ref{FIG_schematics}(b). In particular, the magnetic
field in the $z$ direction is insensitive to the qubit position near
the center between adjacent wires, i.e., odd multiples of $L$. Therefore,
the frequency of the qubit that operated at such locations will be
insensitive to the qubit's displacements along $x$ to the first order.
\citep{2018_SciAdv_Li} The exact sum in Eq. (\ref{Bfield}) can be
carried out, giving the magnetic field at those ideal operation points
$x_{k}\equiv(2k+1)L$ as follows
\begin{equation}
\mathbf{B}(x_{k})=(-1)^{k}B_{0}\zeta\mathrm{sech}\zeta\mathbf{e}_{z},
\end{equation}
where the dimensionless quantity $\zeta=\pi z_{0}/(2L)$. Similarly,
one can also obtain the magnetic field at $x=2kL$ as
\begin{equation}
\mathbf{B}|_{x=2kL}=(-1)^{k}B_{0}\zeta\mathrm{csch}\zeta\mathbf{e}_{x},
\end{equation}
which provides the $B_{x}$ noise insensitive operation points for
the qubit.

If the qubit is located away from $x_{k}$, provided that $z_{0}\ll L$
and $|x-x_{k}|\ll L$ then simplification for the magnetic field at
the qubit location can be still made. This is achieved by restricting
the sum over $n$ in Eq. (\ref{Bfield}) to the nearest two wires
{[}cf. the orange dashed box in FIG. \ref{FIG_schematics}(a){]}.
To ensure the agreement between the overall magnitude for $B$ at
the ideal operation point $x=x_{k}$, one could introduce a calibration
coefficient which we take as $\lambda_{k}=B(x_{k})/[b_{k,1}(x_{k})+b_{k,2}(x_{k})]$,
such that
\begin{equation}
\mathbf{B}(x)\simeq\lambda_{k}\sum_{j=1,2}\mathbf{b}_{k,j}(x),\text{ for }x\simeq x_{k}.\label{Bfield_aprx}
\end{equation}
Here, $\mathbf{b}_{n,j}$ is obtained from Eq. (\ref{SingleWireB})
by replacing $\mathbf{r}_{n}$ and $\mathbf{e}_{n}$ with $\mathbf{r}_{n,j}$
and $\mathbf{e}_{n,j}\equiv(-1)^{j-1}\mathbf{e}_{y}$, respectively.
In FIG. \ref{FIG_schematics}(b), we also show the magnetic field
for $k=0$ calculated from this approximation.

\section{Model and the source of correlated magnetic noises}

In order to implement two-qubit gates in a crossbar architecture,
spins are usually shuttled close to each other such that their mutual
exchange interaction can induce entanglement among the spin qubits.
\citep{2018_SciAdv_Li} In this sense, current fluctuations in a shared
control wire, or a shift of the qubits positions due to mutual charge
interaction, acts as a source of noise in the two-qubit operation
but in a correlated way. Our goals in this section is to provide a
description of such correlated noise at the Hamiltonian level.

The standard model for two-spin 
\begin{figure}
\begin{centering}
\includegraphics[width=8.6cm]{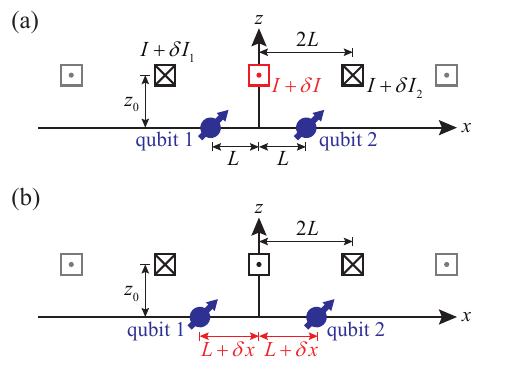}
\par\end{centering}
\caption{Two examples of correlated magnetic noise. (a) Change of the current
at the central wire (highlighted in red) from $I$ to $I+\delta I$
causes related magnetic fields perturbations at the location of qubit
1 and qubit 2. While the independent change of the currents at the
two side wires by $\delta I_{1}$ and $\delta I_{2}$ induces uncorrelated
magnetic field perturbations. (b) Mutual charge repulsion shifts the
qubits locations by $\delta x$ in a correlated way, thus the magnetic
fields at the new locations differs from the original values that
experienced by the qubits. \label{FIG_corrmag}}

\end{figure}
qubits operating in the $(1,1)$ charge configuration is well know,
\citep{1999_PRB_Burkard,2011_PRB_Meunier} which can be described
by the following Hamiltonian upto an overall constant 
\begin{equation}
\hat{H}=J\hat{\mathbf{S}}_{1}\cdot\hat{\mathbf{S}}_{2}+\frac{1}{2}\left[\omega\left(\hat{S}_{1}^{z}+\hat{S}_{2}^{z}\right)+\delta\omega\left(\hat{S}_{1}^{z}-\hat{S}_{2}^{z}\right)\right],\label{Ham}
\end{equation}
where $J$ is the exchange interaction which depends on the tunneling
coupling as well as the detuning between the two spin qubits, $\omega=\omega_{1}+\omega_{2}$
and $\delta\omega=\omega_{1}-\omega_{2}+\delta\omega_{\mathrm{cor}}$
are the sum as well as the difference of local Zeeman splittings $\omega_{1}$
and $\omega_{2}$ at the two qubits locations, while $\delta\omega_{\mathrm{cor}}$
is a correction due to the mixture of different charge states. \citep{Fang2023}

The local Zeeman splittings $\omega_{1}$ and $\omega_{2}$ could
be affected in a correlated way, e.g., when the noisy source acts
on the both qubits. To be more specific, let us consider the configurations
shown in FIG. \ref{FIG_corrmag} where the two spin qubits are shuttled
next to each other in a row along the $x$-axis, typical for implementing
controlled phase operation or spin-to-charge conversion. \citep{2018_SciAdv_Li}
As analyzed in the previous section, we shall focus on the magnetic
fields generated only by the nearest wires {[}cf. FIG. \ref{FIG_schematics}(b){]}.
Thus the Zeeman splittings for the two qubits in the absence of noises
are
\begin{equation}
\omega_{1}=-2\frac{g\mu_{\mathrm{B}}B_{0}}{\hbar}\frac{z_{0}^{2}}{L^{2}+z_{0}^{2}},\text{ }\omega_{2}=2\frac{g\mu_{\mathrm{B}}B_{0}}{\hbar}\frac{z_{0}^{2}}{L^{2}+z_{0}^{2}}.
\end{equation}
Here, $g$ is the g-factor for the spin qubit, $\mu_{\mathrm{B}}$
is the Bohr magneton, and the angle $\theta=\arctan z_{0}/L$. In
general, the perturbation Hamiltonian through the mechanisms shown
in FIG. \ref{FIG_corrmag} could be written into the following form,
\begin{equation}
\delta\hat{H}=\delta\omega_{x}\left(\hat{S}_{1}^{x}+\hat{S}_{2}^{x}\right)-\delta\omega_{z}\left(\hat{S}_{1}^{z}-\hat{S}_{2}^{z}\right),\label{delH_corr}
\end{equation}
while the specific values $\delta\omega_{x}$ and $\delta\omega_{z}$
depend on the details of the noisy scenarios. For the case of FIG.
\ref{FIG_corrmag}(a) with a perturbation $\delta I$ on the central
wire, they are given by
\begin{equation}
\delta\omega_{x}=\frac{g\mu_{\mathrm{B}}\delta B_{0}}{\hbar}\frac{z_{0}^{2}}{L^{2}+z_{0}^{2}},\text{ }\delta\omega_{z}=\frac{g\mu_{\mathrm{B}}\delta B_{0}}{\hbar}\frac{z_{0}L}{L^{2}+z_{0}^{2}},\label{dIB}
\end{equation}
and $\delta B_{0}=\mu_{\mathrm{eff}}\delta I/(2\pi z_{0})$. Instead,
for the position shifts scheme as shown in FIG. \ref{FIG_corrmag}(b),
the values are
\begin{equation}
\delta\omega_{x}=\frac{g\mu_{\mathrm{B}}B_{0}}{\hbar}\left[\frac{z_{0}^{2}}{(L+\delta x)^{2}+z_{0}^{2}}-\frac{z_{0}^{2}}{(L-\delta x)^{2}+z_{0}^{2}}\right],
\end{equation}
\begin{equation}
\delta\omega_{z}=\frac{g\mu_{\mathrm{B}}B_{0}}{\hbar}\left[\frac{z_{0}(L+\delta x)}{(L+\delta x)^{2}+z_{0}^{2}}+\frac{z_{0}(L-\delta x)}{(L-\delta x)^{2}+z_{0}^{2}}\right].
\end{equation}
For comparison, if the current fluctuations $\delta I_{1}$ and $\delta I_{2}$
on the two side wires colored in black in FIG. \ref{FIG_corrmag}(a)
act independently from each other, the resulting perturbation Hamiltonian
would be 
\begin{equation}
\delta\hat{H}=-\delta\omega_{x,1}\hat{S}_{1}^{x}-\delta\omega_{x,2}\hat{S}_{2}^{x}-\delta\omega_{z,1}\hat{S}_{1}^{z}+\delta\omega_{z,2}\hat{S}_{2}^{z},\label{delH}
\end{equation}
where 
\begin{equation}
\delta\omega_{\alpha,i}=\delta\omega_{\alpha}|_{\delta B_{0}=\delta B_{i}},\label{dwi}
\end{equation}
for $i=1,2$ and $\alpha=x,z$, while $\delta B_{i}=\mu_{\mathrm{eff}}\delta I_{i}/(2\pi z_{0})$.
Because here the noise is not due to a correlated source, thus the
perturbation Eq. (\ref{delH}) is formally different from Eq. (\ref{delH_corr}),
expect for special situations, e.g. $\delta I_{1}=\delta I_{2}$ or
$\delta I_{1}=-\delta I_{2}$, where the perturbation Eq. (\ref{delH})
coincide with the form of the correlated perturbation.

\section{Reveal the effects of correlated noises by randomized benchmarking}

In the presence of noises, the actual implementation of a quantum
gate deviates from the ideal one and this usually degrades the gate
fidelity. However, in the context of highly integrated device implementation
with shared controls where the noisy source could affects several
qubits at the same time, it would be more useful to reveal the structure,
e.g., the correlation, of the underlying noises instead of knowing
the gate fidelity alone. 

\begin{figure}
\begin{centering}
\includegraphics[width=8.6cm]{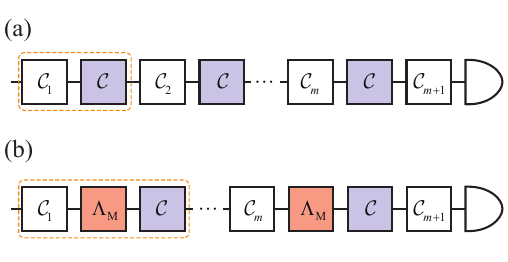}
\par\end{centering}
\caption{(a) Gate sequence in the interleaved banchmakring protocol (IRB),
where $\mathcal{C}_{i}$ is random two-qubit Clifford gate and $\mathcal{C}$
is the gate of interest. After a sequence of $m$ random evolution
unit, the last gate $\mathcal{C}_{m+1}$ reverts the effect of all
previous gates, such that the measurement at the end of the gate sequence
yields a result corresponding to the initial state with unit probability
in the absence of noisy effects. (b) The modified IRB protocol, where
a measurement process described with the quantum channel $\Lambda_{\mathrm{M}}$
is applied after each random Clifford gate. \label{FIG3_ModifiedIRB}}
\end{figure}

In this section, we investigate the correlated noisy effects of the
model configuration in FIG. \ref{FIG_corrmag}(a) by the interleaved
randomized benchmarking (IRB). \citep{2012_PRL_Magesan} Our scheme
is illustrated in FIG. \ref{FIG3_ModifiedIRB}, where 
 in additional to the standard circuit implementing the IRB. We introduced
a measurement process $\Lambda_{\mathrm{M}}$ in between the random
Clifford gates and the gate of interest $\hat{U}_{\mathrm{c}}$. By
chosen the basis for the measurement process, certain components of
the noisy channel could be selected, such that one can distinguish
between the two particular cases $\delta I_{1}=\delta I_{2}$ and
$\delta I_{1}=-\delta I_{2}$ based on the scenario of FIG. \ref{FIG_corrmag}(a),
here we refer them as the \emph{correlated} and the \emph{anti-correlated}
noisy perturbations.

For simplicity, we shall assume that in the ideal case $\hat{U}_{\mathrm{c}}$
is also an element from the two-qubit Clifford group $\mathcal{C}_{2}$.
This can be achieved from time evolution of the Hamiltonian $\hat{H}$
with the following conditions: \citep{Fang2023}
\begin{equation}
t=\pi J^{-1},\text{ }\delta\omega=\sqrt{15}J,\text{ }\omega=4kJ,
\end{equation}
where $k$ is any positive integer. With the above conditions, in
the computational basis one has
\begin{equation}
\hat{U}_{\mathrm{c}}=e^{-i\hat{H}t}=e^{-i\pi/4}\left[\begin{array}{cccc}
1 & 0 & 0 & 0\\
0 & i & 0 & 0\\
0 & 0 & i & 0\\
0 & 0 & 0 & 1
\end{array}\right],
\end{equation}
which belongs to $\mathcal{C}_{2}$ up to an overall phase factor.

The actually implemented gate of interest (denoted as $\hat{U}$)
evolves from the full Hamiltonian $\hat{H}+\delta\hat{H}$ and, in
the presence of magnetic noises, it
\begin{figure*}
\begin{centering}
\includegraphics[width=17.2cm]{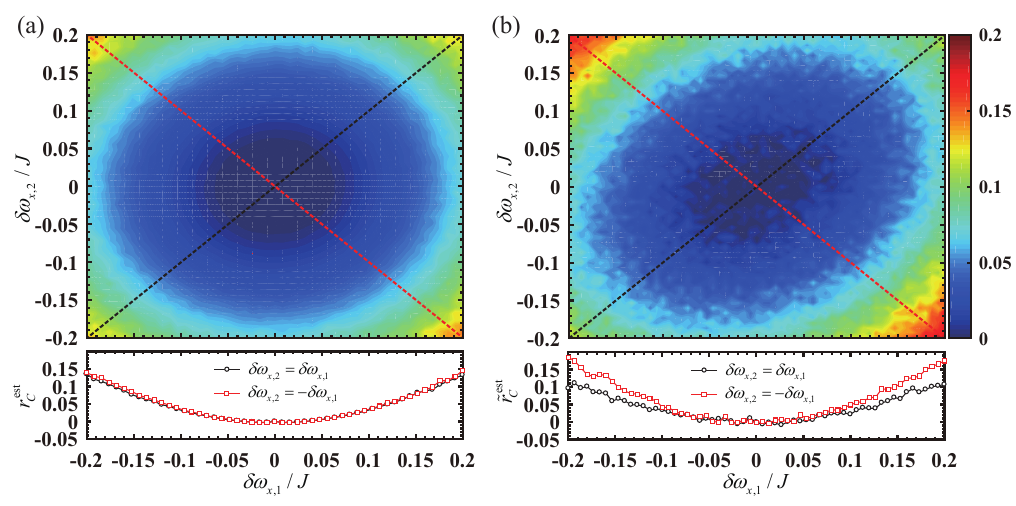}
\par\end{centering}
\caption{(a) Main panel: average error rate $r_{C}^{\mathrm{est}}$associated
with the actual implementation of the two-qubit gate $\hat{U}_{\mathrm{c}}$
subjected to independent noises $\delta I_{1}$ and $\delta I_{2}$,
which are proportional to the Zeeman splittings according to Eqs.(\ref{dIB},\ref{dwi}).
The error rate is obtained from the standard IRB protocol. Lower panel:
$r_{C}^{\mathrm{est}}$ along the black and red dashed cuts in the
main panel. (b) Main panel: The error rate $\tilde{r}_{C}^{\mathrm{est}}$
extracted from the modified protocol as in FIG. \ref{FIG3_ModifiedIRB}(b),
lower panel: $\tilde{r}_{C}^{\mathrm{est}}$ along the black and red
cuts of the main panel. Here, we choose the initial state $|\uparrow,\uparrow\rangle$
for simulation, and $\hat{P}=|\mathrm{T}_{0}(1,1)\rangle\langle\mathrm{T}_{0}(1,1)|$
for constructing $\Lambda_{\mathrm{M}}$. Parameters used in the calculation:
$t=\pi/J$, $\omega=20J$, $\delta\omega=\sqrt{15}J$, $z_{0}=L$,
$N_{\mathrm{avg}}=10^{3}$, $m=200$ in (a) and $m=30$ in (b). \label{FIG_IRBresult}}
\end{figure*}
 deviates from $\hat{U}_{\mathrm{c}}$. The associated effect of the
gate error on time evolution can be formally written in terms of a
quantum channel $\Lambda_{\mathrm{c}}$ given by
\begin{equation}
\Lambda_{\mathrm{c}}[\hat{\rho}]=\delta\hat{U}\hat{\rho}\delta\hat{U}^{\dagger},
\end{equation}
where $\delta\hat{U}=\hat{U}_{\mathrm{c}}^{\dagger}\hat{U}$. Thus
the time evolution of the actual gate is written as $\hat{U}\hat{\rho}\hat{U}^{\dagger}=\mathcal{C}\circ\Lambda_{\mathrm{c}}[\hat{\rho}]$
with $\mathcal{C}[\hat{\rho}]=\hat{U}_{\mathrm{c}}\hat{\rho}\hat{U}_{\mathrm{c}}^{\dagger}$. 

Following the standard approach (see also Appendix A for a brief review),
\citep{2012_PRA_Magesan,2012_PRL_Magesan} we prepared the two-qubit
system in the initial state $|\uparrow,\uparrow\rangle$ and performed
the usual randomized benchmarking (RB) to extract the depolarization
parameter $p$ associated with noise channel $\Lambda$ for the random
Clifford gates. For simplicity, we assume a gate independent noise
channel $\Lambda$ such that the zero-th order fitting model is applicable
to extract $p$. Then by interleaving the actual implementation of
$\hat{U}_{\mathrm{c}}$ in between each random Clifford gate {[}cf.
FIG. \ref{FIG3_ModifiedIRB}(a){]}, one obtains a modified depolarization
parameter $p_{\mathrm{c}}$ that associates with the combined noisy
channel $\Lambda_{\mathrm{c}}\circ\Lambda$. From $p_{\mathrm{c}}$
and $p$ the average error rate of $\Lambda_{\mathrm{c}}$ could be
estimated from \citep{2012_PRL_Magesan}
\begin{equation}
r_{C}^{\mathrm{est}}=\frac{d-1}{d}\left(1-\frac{p_{\mathrm{c}}}{p}\right),\label{rcest}
\end{equation}
where $d$ is the dimension of the Hilbert space. In the absence of
noises for implementing the random Clifford gates, $p=1$ and Eq.
(\ref{rcest}) reduces to the well known RB result $(d-1)(1-p_{\mathrm{c}})/d$.\citep{2011_PRL_Magesan,2012_PRA_Magesan}

From the above approach, the average error rate for the actual implementation
of the gate $\hat{U}_{\mathrm{c}}$ is simulated and shown in FIG.
\ref{FIG_IRBresult}(a). The result suggests that $r_{C}^{\mathrm{est}}$
obtained from the standard IRB is symmetric with respective to rotations
in the $\delta I_{1}$-$\delta I_{2}$ plane, i.e., $r_{C}^{\mathrm{est}}$
depends only on $\sqrt{\delta I_{1}^{2}+\delta I_{2}^{2}}$. This
is more clearly seen by comparing the two special cases with $\delta I_{1}=\delta I_{2}$
and $\delta I_{1}=-\delta I_{2}$ as shown in the lower panel of FIG.
\ref{FIG_IRBresult}(a), where the calculated $r_{C}^{\mathrm{est}}$
overlaps with each other, even though those two cases represent completely
different correlations among the perturbations applied to the side
wires {[}cf. FIG. \ref{FIG_corrmag}(a){]}. 

To better distinguish between the \emph{correlated} and the \emph{anti-correlated}
cases, we modify the gate sequence in the standard IRB protocol, which
is illustrated in FIG. \ref{FIG3_ModifiedIRB}(b). As mentioned before,
the main difference here is to introduce a measurement induced decoherence
process, given in term of a noisy channel $\Lambda_{\mathrm{M}}[\hat{\rho}]=\hat{P}\hat{\rho}\hat{P}^{\dagger}+\hat{Q}\hat{\rho}\hat{Q}^{\dagger}$,
where $\hat{P}$ and $\hat{Q}$ are complementary projection operators
satisfying $\hat{P}+\hat{Q}=\hat{I}$. Introducing $\Lambda_{\mathrm{M}}$
is motivated by the fact that the insensitive of the standard IRB
circuit to $\delta I_{1}$ and $\delta I_{2}$ could be related to
the symmetry of the perturbation Hamiltonian. In particular, one can
show that for the perturbation term $\delta\hat{H}$ given by Eq.
(\ref{delH}), there are two dark states for each of the two special
cases, i.e., $\delta\hat{H}_{\lambda}|D_{i}\rangle_{\lambda}=0$ with
$\lambda\in\{\mathrm{S},\mathrm{A}\}$, where $\delta\hat{H}_{\lambda}$
is defined as follows
\begin{equation}
\delta\hat{H}_{\mathrm{S}}=\delta\hat{H}|_{\delta I_{1}=\delta I_{2}},\text{ }\delta\hat{H}_{\mathrm{A}}=\delta\hat{H}_{\delta I_{1}=-\delta I_{2}}.\label{dHA_dHS}
\end{equation}
The specific form of those dark states are listed in the Appendix
B. Furthermore, the projectors $\hat{P}_{\lambda,i}=|D_{i}\rangle_{\lambda}\langle D_{i}|$
satisfy $\hat{P}_{\mathrm{S},i}\hat{P}_{\mathrm{A},j}=0$. Therefore,
projection with $\hat{P}_{\mathrm{S},i}$ ($\hat{P}_{\mathrm{A},i}$)
during the implementation of the random circuit will reduce the effects
of noisy perturbation in the \emph{anti-correlated} (\emph{correlated})
case and lead to higher sequence fidelity, thus enables the distinguishing
between the two situations.

Based on the scheme of FIG. \ref{FIG3_ModifiedIRB}(b), the sequence
fidelity from a particular realization of the random circuit becomes
\begin{equation}
F_{\mathrm{seq}}=\mathrm{Tr}\left\{ \hat{E}_{\xi}\Lambda\circ\mathcal{C}_{m+1}\left[\prod_{i=1}^{m}\mathcal{C}\circ\Lambda_{\mathrm{c}}\circ\Lambda_{\mathrm{M}}\circ\Lambda\circ\mathcal{C}_{i}\right]\hat{\rho}_{0}\right\} ,\label{Fseq_new}
\end{equation}
where $\mathcal{C}_{m+1}=[\prod_{i=1}^{m}\mathcal{C}\circ\mathcal{C}_{i}]^{-1}$.
Then by averaging $F_{\mathrm{seq}}$ over $N_{\mathrm{avg}}$ realizations
over the random circuits, one has
\begin{equation}
\langle F_{\mathrm{seq}}\rangle=\mathrm{Tr}\left\{ \hat{E}_{\xi}\Lambda\circ[\Lambda_{\mathrm{c}}\circ\Lambda_{\mathrm{M}}\circ\Lambda]_{\mathrm{d}}^{m}\hat{\rho}_{0}\right\} \equiv A\tilde{p}_{\mathrm{c}}^{m}+B,
\end{equation}
where $\tilde{p}_{\mathrm{c}}$ is the depolarization parameter associated
with the combined depolarization noisy channels $[\Lambda_{\mathrm{c}}\circ\Lambda_{\mathrm{M}}\circ\Lambda]_{\mathrm{d}}$,
after taking the twirling operation over the two-qubit Clifford group.
\citep{2009_PRA_Dankert,2012_PRA_Magesan} Alternatively, removing
$\mathcal{C}\circ\Lambda_{\mathrm{c}}$ from Eq. (\ref{Fseq_new})
provides a circuit that gives rise to a referential value for the
sequence fidelity, whose average leads to the depolarization parameter
$\tilde{p}$ associated with $[\Lambda_{\mathrm{M}}\circ\Lambda]_{\mathrm{d}}$.
Thus by inserting $\tilde{p}$ and $\tilde{p}_{\mathrm{c}}$ into
Eq. (\ref{rcest}), one can extract another error rate $\tilde{r}_{C}^{\mathrm{est}}$
associated with gate of interest. 

With properly chosen basis for the measurement related projection
operator $\hat{P}$, the error rate $\tilde{r}_{C}^{\mathrm{est}}$
could distinguish between the \emph{correlated} and \emph{anti-correlated}
noises cases. This is shown in FIG. \ref{FIG_IRBresult}(b) for $\hat{P}=|\mathrm{T}_{0}(1,1)\rangle\langle\mathrm{T}_{0}(1,1)|$,
where $|\mathrm{T}_{0}(1,1)\rangle$ is the triplet state $|\uparrow,\downarrow\rangle+|\downarrow,\uparrow\rangle$
and also one of the dark states for $\delta\hat{H}_{\mathrm{A}}$.
Here, the error rate shows asymmetry as $\delta I_{1}$ and $\delta I_{2}$
varying, and the error rate grows more slowly for the \emph{correlated}
noise perturbation as $\delta I_{1}$ and $\delta I_{2}$ increasing.

\section{Discussions}

Although here we only analysis through a particular perturbation $\delta\hat{H}$
that described by the configuration of FIG. \ref{FIG_corrmag}(a),
the idea of introducing measurement induced decoherence to distinguish
noisy correlations should be also applicable to other types of noisy
sources, for example, apart from alternating current-carrying wires,
other way of generating the magnetic field gradient include the use
of micromagnets,\citep{2006_PRL_Tokura,2008_NatPhys_Pioro} as well
as the Overhauser fields from polarized nuclear spins.\citep{2007_PRL_Baugh,2008_PSSC_Baugh}
However, in order to construct $\tilde{r}_{C}^{\mathrm{est}}$ that
will be sensitive to the noisy correlation, the proper choice of projection
operators $\hat{P}$ and $\hat{Q}$ are crucial. To find out those
operators, one may wish to enable simpler calculation instead of the
full IRB simulations. In fact, we found similar asymmetry behavior
in the depolarization parameter $\tilde{p}_{\mathrm{c}}$ of $[\Lambda_{\mathrm{c}}\circ\Lambda_{\mathrm{M}}\circ\Lambda]_{\mathrm{d}}$.
Notice that this is directly related to the twirling of $\Lambda_{\mathrm{c}}\circ\Lambda_{\mathrm{M}}\circ\Lambda$
under the two-qubit Clifford group. Thus for theoretical analysis,
it is useful to first consider $[\Lambda_{\mathrm{c}}\circ\Lambda_{\mathrm{M}}\circ\Lambda]_{\mathrm{d}}$
instead of $\tilde{r}_{C}^{\mathrm{est}}$.

The sequence fidelity depends also on the system initial state, thus
choosing proper initial state could be also useful to distinguish
the noisy correlations. For example, suppose the initial state of
the two-qubit system is $|\mathrm{T}_{0}(1,1)\rangle\propto|\uparrow,\downarrow\rangle+|\downarrow,\uparrow\rangle$,
since this state is also a dark state of $\delta\hat{H}_{\mathrm{A}}$
{[}cf. Eq. (\ref{A2_DS3}){]}, thus calculating with this initial
state amounts to applying the projection operator $\hat{P}$ at the
beginning and end of the random circuit. Throughout the calculation
we have assumed that the two-qubit system is prepared in the $|\uparrow,\uparrow\rangle$
state. For spin qubits based on the double quantum dot, this state
can be prepared by adiabatically changing the detuning through the
$\mathrm{S}$-$\mathrm{T}_{+}$ anti-crossing that induced with the
transverse field of micromagnet. \citep{2014_PNAS_Wu,2014_PRB_Chesi} 

For the purpose of illustration, we did not apply realistic parameters
in the simulation. In silicon based double quantum dot, the exchange
coupling above $10\mathrm{MHz}$ had been measured,\citep{2015_Nature_Veldhorst}
while for the crossbar system, the magnetic field was designed such
that $\delta\omega\sim10\mathrm{MHz}$ and $\omega\sim100\mathrm{MHz}$.
\citep{2018_SciAdv_Li} Also, the calculations were performed in the
absence of qubit relaxation or dephasing, and the dynamical aspect
of the current fluctuation in $\delta I_{1}$ and $\delta I_{2}$
are ignored. The detailed treatment of those effects are out of the
current scope and will be addressed in the future works.

\section{Conclusions}

In this paper, we have explored correlated magnetic noise in a two-qubit
model stimulated by the recent proposals of scalable quantum computation
in semi-conductor based spin qubits. With the method introduced here,
we could distinguish between the \emph{correlated} and \emph{anti-correlated}
noisy cases from their different magnetic field dependence of the error
rate, obtained from the interleaved randomized bechmarking protocol
by using the modified random circuit. The key idea is to introduce
a measurement based docherence process such that certain component
of the noisy channel associated with the correlated noise could be
singled out.
\begin{acknowledgments}
Y.F. acknowledges support from NSFC (Grant No. 12005011) and Yunnan
Fundamental Research Projects (Grant No. 202201AU070118).
\end{acknowledgments}

\appendix

\section{Interleaved randomized benchmarking}

In this Appendix, we follow Refs. \citep{2005_JPhysB_Emerson,2011_PRL_Magesan,2012_PRL_Magesan,2012_PRA_Magesan}
to provide a brief summary of the interleaved randomized benchmarking
(IRB) protocol.

In general, the protocol involves three steps: First perform the usual
randomized benchmarking (RB) to extract $p$;\citep{2005_JPhysB_Emerson,2011_PRL_Magesan,2012_PRA_Magesan}
Then perform an interleaved version of the random circuits to extract
$p_{\mathrm{c}}$;\citep{2012_PRL_Magesan} Finally, from $p_{\mathrm{c}}$
and $p$ one obtain an estimate for the average error rate $r_{C}^{\mathrm{est}}$
associated with the gate of interest, given in terms of the average
fidelity $r_{C}^{\mathrm{est}}=1-F_{\mathrm{avg}}[\Lambda_{\mathrm{c}}]$
associated with noisy channel to implement the gate of interest
\begin{equation}
F_{\mathrm{avg}}[\Lambda_{\mathrm{c}}]=\int d\psi\langle\psi|\Lambda_{\mathrm{c}}[|\psi\rangle\langle\psi|]|\psi\rangle.\label{A1_Favg}
\end{equation}

Next, we describe the above steps in more detail. In the first step,
to extract $p$ the sequence fidelity is calculated as a function
of the number of random gates $m$ applied to the circuit
\begin{equation}
F_{\mathrm{seq}}=\mathrm{Tr}\left\{ \hat{E}_{\xi}\Lambda\circ\mathcal{C}_{m+1}\left[\prod_{i=1}^{m}\Lambda\circ\mathcal{C}_{i}\right]\hat{\rho}_{0}\right\} ,
\end{equation}
where $\hat{\rho}_{0}$ is the initial state, $\hat{E}_{\xi}$ is
the final measurement operator. In the absence of state preparing
and measurement errors one could set $\hat{\rho}_{0}=|\psi_{0}\rangle\langle\psi_{0}|$
and also $\hat{E}_{\xi}=|\psi_{0}\rangle\langle\psi_{0}|$. $\mathcal{C}_{i}[\hat{\rho}]=\hat{U}(g_{i})\hat{\rho}\hat{U}(g_{i})^{\dagger}$
with $g_{i}\in\mathcal{C}_{n}$ being an element of the $n$-qubit
Clifford group\noun{ $\mathcal{C}_{n}$}, while $\Lambda$ is the
gate-independent quantum channel describing the noisy effect involved
in implementing the Clifford gate $\mathcal{C}_{i}$. Notice that
there are only $m$ random gates, which uniquely determines the last
gate $\mathcal{C}_{m+1}\equiv[\mathcal{C}_{m}\circ...\circ\mathcal{C}_{1}]^{-1}$.
Since $F_{\mathrm{seq}}$ is random because of $\{\mathcal{C}_{i}\}$.
Thus one should consider the averaged quantity $\langle F_{\mathrm{seq}}\rangle$
by random sampling over the random Clifford gates, for sufficiently
large number of samples $N_{\mathrm{avg}}$, the averaged sequence
fidelity approximates the following exact value
\begin{equation}
\langle F_{\mathrm{seq}}\rangle\overset{\mathrm{large}N_{\mathrm{avg}}}{\simeq}\frac{1}{|\mathcal{C}_{n}|^{m}}\sum_{\mathcal{C}_{1}..\mathcal{C}_{m}}F_{\mathrm{seq}}.
\end{equation}
Since the Clifford group form a unitary 2-design,\citep{2009_PRA_Dankert}
thus the discrete sum over the Clifford group element can reproduce
the continuous average over $\mathrm{SU}(n)$ appeared in the definition
of the average fidelity Eq. (\ref{A1_Favg}). In particular, the twirling
operation on $\Lambda$ by the Clifford group yields a depolarization
channel
\begin{equation}
[\Lambda]_{\mathrm{tw}}\equiv\frac{1}{|\mathcal{C}_{n}|}\sum_{g\in\mathcal{C}_{n}}\mathcal{C}(g)\circ\Lambda\circ\mathcal{C}(g)^{-1}=[\Lambda]_{\mathrm{d}},
\end{equation}
such that $[\Lambda]_{\mathrm{d}}\hat{\rho}=p\hat{\rho}+(1-p)\hat{I}/d$,
where $p$ is known as the depolarization parameter. Importantly,
due to the unitary 2-design property, the average fidelity of $[\Lambda]_{\mathrm{d}}$
is the same as that for $\Lambda$.\citep{2012_PRL_Magesan} By the
group rearrangement theorem, $[\Lambda]_{\mathrm{d}}$ commutes with
arbitrary Clifford gate $\mathcal{C}(g)$, thus the average sequence
fidelity follows simply the zero-th order fitting model given by \citep{2011_PRL_Magesan}
\begin{equation}
\langle F_{\mathrm{seq}}\rangle=Ap^{m}+B.\label{A1_fitmodel}
\end{equation}

In the second step, to extract $p_{\mathrm{c}}$ one interleaves the
gate of interest $\hat{U}_{\mathrm{c}}$ in between random Clifford
gates, thus the sequence fidelity of the modified circuit becomes
\begin{equation}
F_{\mathrm{seq,c}}=\mathrm{Tr}\left\{ \hat{E}_{\xi}\Lambda\circ\mathcal{C}_{m+1}\left[\prod_{i=1}^{m}\mathcal{C}\circ\Lambda_{\mathrm{c}}\circ\Lambda\circ\mathcal{C}_{i}\right]\hat{\rho}_{0}\right\} ,
\end{equation}
where $\mathcal{C}[\hat{\rho}]=\hat{U}_{\mathrm{c}}\hat{\rho}\hat{U}_{\mathrm{c}}^{\dagger}$
and $\mathcal{C}_{m+1}=[\prod_{i=1}^{m}\mathcal{C}\circ\mathcal{C}_{i}]^{-1}$.
Similar consideration leads to the conclusion that the average sequence
fidelity again follows the fitting model Eq. (\ref{A1_fitmodel})
but with modified parameters
\begin{equation}
\langle F_{\mathrm{seq,c}}\rangle=Ap_{\mathrm{c}}^{m}+B.
\end{equation}
Now $p_{\mathrm{c}}$ is the depolarization parameter of $[\Lambda_{\mathrm{c}}\circ\Lambda]_{\mathrm{d}}$. 

In the third step, one estimates $r_{C}^{\mathrm{est}}$ according
to Eq. (\ref{rcest}), which can be derived by identifying the average
fidelity of $\Lambda_{\mathrm{c}}\circ\Lambda$ with that of $\Lambda_{\mathrm{c}}\circ[\Lambda]_{\mathrm{d}}$.\citep{2012_PRL_Magesan}

\section{Dark states of $\delta\hat{H}$}

Here, we list all the dark states for $\delta\hat{H}_{\mathrm{S}}$
and $\delta\hat{H}_{\mathrm{A}}$, i.e., the perturbation term defined
in Eq. (\ref{dHA_dHS}). For $\delta\hat{H}_{\mathrm{S}}$ the two
dark states are
\begin{equation}
|D_{1}\rangle_{\mathrm{S}}=\frac{1}{\sqrt{2}}\left[|\uparrow,\uparrow\rangle-|\downarrow,\downarrow\rangle\right],\label{A2_DS1}
\end{equation}
and
\begin{equation}
|D_{2}\rangle_{\mathrm{S}}=\frac{1}{2}\left[|\uparrow,\uparrow\rangle-|\uparrow,\downarrow\rangle+|\downarrow,\uparrow\rangle+|\downarrow,\downarrow\rangle\right].
\end{equation}
While for $\delta\hat{H}_{\mathrm{A}}$ the two dark states are
\begin{equation}
|D_{1}\rangle_{\mathrm{A}}=\frac{1}{\sqrt{2}}\left[|\uparrow,\downarrow\rangle+|\downarrow,\uparrow\rangle\right],\label{A2_DS3}
\end{equation}
and
\begin{equation}
|D_{2}\rangle_{\mathrm{A}}=\frac{1}{2}\left[|\uparrow,\uparrow\rangle+|\uparrow,\downarrow\rangle-|\downarrow,\uparrow\rangle+|\downarrow,\downarrow\rangle\right].
\end{equation}

\bibliographystyle{apsrev4-1}
\bibliography{References}

\end{document}